\documentclass{aastex62}

\usepackage[utf8]{inputenc}
\usepackage[english]{babel}
\usepackage{amsmath}
\usepackage{amsfonts}
\usepackage{amssymb}
\usepackage{graphicx}
\usepackage{gensymb}
\usepackage{tabularx}
\usepackage{array}
\usepackage{natbib}
\setcitestyle{round}
\usepackage{hyperref}
\newcolumntype{Y}{>{\centering\arraybackslash}X}

\providecommand{\keywords}[1]
{
  \small	
  \textbf{\textit{Keywords---}} #1
}

\RequirePackage{natbib}
\renewcommand\cite{\citep}

\begin{document}
%\onecolumn[
%  \begin{@twocolumnfalse}
%    \maketitle

\title{\Large Crash Chronicles: relative contribution from comets and carbonaceous asteroids to Earth's volatile budget in the context of an \textit{Early Instability}}

\author[0000-0001-5985-2863]{Sarah Joiret}
\affiliation{Laboratoire d'Astrophysique de Bordeaux, Univ. Bordeaux, CNRS \\ 
B18N, allée Geoffroy Saint-Hilaire \\
33615 Pessac, France}
\author{Sean N. Raymond}
\affiliation{Laboratoire d'Astrophysique de Bordeaux, Univ. Bordeaux, CNRS \\ 
B18N, allée Geoffroy Saint-Hilaire \\
33615 Pessac, France}
\author{Guillaume Avice}
\affiliation{Université Paris Cité, Institut de physique du globe de Paris, CNRS \\
75005 Paris, France}
\author{Matthew S. Clement}
\affiliation{Johns Hopkins APL, 11100 Johns Hopkins Rd \\ 
Laurel, MD 20723, USA}

    \begin{abstract}
Recent models of solar system formation suggest that a dynamical instability among the giant planets happened within the first 100 Myr after disk dispersal, perhaps before the Moon-forming impact. As a direct consequence, a bombardment of volatile-rich impactors may have taken place on Earth before internal and atmospheric reservoirs were decoupled. However, such a timing has been interpreted to potentially be at odds with the disparate inventories of Xe isotopes in Earth's mantle compared to its atmosphere. This study aims to assess the dynamical effects of an \textit{Early Instability} on the delivery of carbonaceous asteroids and comets to Earth, and address the implications for the Earth's volatile budget. We perform 20 high-resolution dynamical simulations of solar system formation from the time of gas disk dispersal, each starting with 1600 carbonaceous asteroids and 10000 comets, taking into account the dynamical perturbations from an early giant planet instability. Before the Moon-forming impact, the cumulative collision rate of comets with Earth is about 4 orders of magnitude lower than that of carbonaceous asteroids.  After the Moon-forming impact, this ratio either decreases or increases, often by orders of magnitude, depending on the dynamics of individual simulations. An increase in the relative contribution of comets happens in 30\% of our simulations. In these cases, the delivery of noble gases from each source is comparable, given that the abundance of $^{132}Xe$ is 3 orders of magnitude greater in comets than in carbonaceous chondrites. The increase in cometary flux relative to carbonaceous asteroids at late times may thus offer an explanation for the Xe signature dichotomy between the Earth’s mantle and atmosphere. Our current model falls short in simultaneously reproducing a 4 - 10 \% carbonaceous contribution to Earth and late accretion dominated by non-carbonaceous material.  In future work, it is worth exploring the possibility that a large fraction of Earth's carbonaceous material was accreted prior to gas disk dispersal.  If  that were the case, it would strengthen our dynamical explanation for the Xe signature dichotomy, because the contribution of cometary relative to carbonaceous material would strongly increase at later times.

    \end{abstract}
    \vspace{4mm}
%  \end{@twocolumnfalse}
%]

\section{Introduction}
\label{Intro}
The origin of volatile elements on Earth is closely tied to the early dynamical evolution of the solar system. A comparison between the isotopic signatures of chondritic meteorites and those of Earth suggest that Earth and the other terrestrial planets mostly formed from building blocks native to the inner solar system \cite{Dauphas2017, Burkhardt2021}.  While this locally-sourced material likely provided a significant amount of Earth's water \cite{Piani2020}, a volatile-rich contribution from the outer Solar System is needed to fully explain Earth's volatile budget \cite{Morbidelli2000, Raymond2004, Marty2012, Marty2016}. This contribution from carbonaceous asteroids and comets probably took the form of a bombardment, as a result of the giant planets' growth and migration while the gas disk was still present \cite{Walsh2012, Raymond2017}  and a later dynamical evolution in the form of a dynamical instability \cite{Gomes2005}. However, the timing of this giant planet instability has recently been revised and this has implications for the Earth's volatile budget. 

\subsection{Giant planet dynamical instability and its timing}
It is commonly agreed that dynamical instabilities are widespread among other planetary systems as they could explain the surprisingly large eccentricities of most giant exoplanets \cite{Rasio1996, Weidenschilling1996, Lin1997, Adams2003, Ford2003, Juric2008, Chatterjee2008, Raymond2010}. In the case of the solar system, there are several observational constraints that support the instability hypothesis \cite{Nesvorny2018review} including the eccentricity of the giant planets \cite{Tsiganis2005}, the orbital structure of the asteroid and Kuiper belts \cite{Hahn2005, Morbidelli2010, Nesvorny2015, Deienno2016, Deienno2018, Volk2019, Clement2020, Nesvorny2021} and the cratering history on the Moon and terrestrial planets \cite{Bottke2017}.

Initially, the instability was invoked as a late event \cite{Gomes2005} based mainly on radiometric ages obtained on lunar samples \cite{Tera}, but recent re-analysis of constraints suggests that it instead happened within the first 100 Myr of the solar system \cite{Boehnke2016, Zellner, Nesvorny2018, Morby2018, Mojzsis2019, Hartmann2019}, and even possibly shortly after the dispersal of the gas disk \cite{Ribeiro2020, Liu2022}. 

The timing of the giant planets' dynamical instability depends on what dynamical mechanism served as a trigger.  To date, four candidates have been identified as possible triggers: a) direct gravitational interaction between the planets and an outer planetesimal disk  \cite{Tsiganis2005, Morby2007, Batygin2011, Ribeiro2020}; b) indirect gravitational interaction between the planets and an outer, self-gravitating planetesimal disk \cite{Levison2011, Quarles19}; c) self-driven instability \cite{Ribeiro2020}, i.e. giant planets' orbits go unstable on their own, with no appreciable interaction with the planetesimal disk; and d) differential migration driven by the inside-out dispersal of the gaseous protoplanetary disk \cite{Liu2022}.  Mechanisms c and d have each been shown to generally trigger rapid dynamical instability within at most a few Myr of the dispersal of the gaseous protoplanetary disk \cite{Ribeiro2020, Liu2022}.  Interactions between the planets and a consistently-sculpted planetesimal disk (mechanism a) tend to induce dynamical instability on a timescale of $\simeq$ 50 Myr \cite{Deienno2017, Ribeiro2020}.  Taking into account the self-gravity of the disk (mechanism b) tends to significantly shorten the time to instability \cite{Quarles19} (see \citet{Levison2011} for an alternative view), although simulations that account for both the early dynamical sculpting of the belt and its self-gravity have never been run to our knowledge.

In this paper we assume that the dynamical instability took place early, meaning within the first $\simeq$ 10-20 Myr after dispersal of the gaseous disk. \citet{Clement2018} showed that an instability within the first 10 Myr of the solar system is able to replicate the orbits and masses of the terrestrial planets - including the small mass of Mars \cite{Raymond2009} - through numerical simulations. The so-called \textit{Early Instability} model is corroborated by thermochronologic data recorded in meteorites \cite{Edwards2023}. 

\subsection{Delivery of volatiles before and after the Moon-forming impact}
The Moon-forming impact happened between $\simeq$ 30 to 200 Myr after the solar system was born, as measured by Hf/W and U/Pb systematics \cite{Kleine, Rudge, Maurice2020}. A bombardment of asteroids and comets occurring before this last giant impact has implications for the delivery of volatiles to the Earth's mantle. 

A signature of early volatile-rich impactors is expected to be retained in both Earth's mantle and atmosphere, whereas later impactors mostly left their volatile imprint within Earth's atmosphere due to decoupling of internal and atmospheric reservoirs after magma ocean solidification. Between the dissipation of the primary atmosphere and the Moon-forming impact, Earth's atmosphere was oftentimes in a runaway greenhouse state interacting with a magma ocean surface \cite{Abe1985, Abe1986, Zahnle1988, Zahnle2007, Hamano2013, Young2023}. While degassing of volatile-rich planetesimals upon impacts may have imparted most of their volatile budget directly into the atmosphere \cite{Benlow1977}, partial volatile retention in the solid or melted portion of the mantle is also likely \cite{Davison2008, Elkins2008}. More generally, volatile exchange between the Earth's mantle and atmosphere has been facilitated by liquid–gas interactions at the surface. Throughout magma ocean solidification, all volatiles in excess of the saturation limits of the silicate liquid would have been degassed into the growing atmosphere \cite{Zahnle1988, Elkins2008}.  

\subsection{Late accretion}
The mass accreted to Earth after the last giant impact, i.e. the Moon-forming impact, is referred as the late accretion. This mass is often constrained from the concentration of highly siderophile elements (HSEs) in Earth’s mantle \cite{Chyba1991, Walker2009, Bottke2010}, as these elements should have been partitioned into the core if they had been delivered prior to Earth's differentiation \cite{Kimura1974}. Using dynamical simulations, \citet{Jacobson} found that the late accretion mass is inversely correlated with the time of the Moon-forming impact. Assuming that the late accretion mass is $\simeq$ 0.5\% of the Earth's final mass, they argued that the time of the Moon-forming impact occurred significantly later than 40 Myr after condensation of the first solids in the solar system. It should be noted, however, that the abundance of HSEs in the mantle depends on the degree of equilibration between metal and silicate after each impact, which in turn depends on whether the impactors were large, differentiated bodies or not \cite{Deguen2011, Morby2015}.

\subsection{Chondritic and cometary geochemical constraints}
Nucleosynthetic isotope anomalies are used to distinguish between different precursor materials on Earth, in particular non-carbonaceous and carbonaceous sources. Indeed, an isotopic dichotomy was identified between non-carbonaceous (NC) and carbonaceous (CC) meteorites from anomalies in Ti, Cr, Ni, and Mo \cite{Trinquier2007, Trinquier2009, Warren2011, Kleine2020}. In this manner, and depending on the subset of elements considered, the contribution of carbonaceous asteroids to the Earth's mantle has recently been limited to a few percent of Earth by mass. It is close to 4\% \cite{Burkhardt2021} or 6\% \cite{Savage2022} if we consider CI chondrite-like material, or at most 10\% if we consider CV chondrite-like material \cite{Kleine2023}, with the remaining mass coming from non-carbonaceous bodies. Even though carbonaceous material accounts for a relatively minor component of Earth's total mass, it likely delivered a significant amount ($\simeq$ 30\%) of Earth’s budget of moderately volatile elements \cite{Savage2022, Steller2022}. 

Cometary isotopic ratios are either obtained remotely by high-resolution spectroscopy or with space-based observations. For decades, the distinction between the two types of volatile-rich precursors (i.e. comets and carbonaceous asteroids) could mostly be made by comparing their respective D/H \cite{Hartogh2011, Ceccarelli2014} and $^{15}N$/$^{14}N$ ratios \cite{Jehin2009, Manfroid2009}. More recently, ESA's Rosetta spacecraft was able to measure the noble gas content and isotope ratios in the coma of comet 67P/Churyumov-Gerasimenko \cite{Balsiger2007, Balsinger2015, Rubin2018}. Using the full data set, it was found that while the contribution of cometary volatiles to the Earth's budget is minor for water ($\le$ 1\%), carbon ($\le$ 1\%), and nitrogen species (a few \% at most), it is significant for the atmospheric noble gases \cite{Marty2012, Marty2016}. This can be explained by the fact that the noble gases to water abundance ratios (Ar/$H_{2}0$, Kr/$H_{2}0$ and Xe/$H_{2}0$) are several orders of magnitude larger in comets than in chondrites \cite{Balsinger2015, Bekaert2020}. This implies that, when comets bombarded the Earth, they could have largely contributed to the noble gas inventory without altering significantly the D/H ratio \cite{Marty2012}, not to mention some comets might actually have D/H ratios similar to those of carbonaceous chondrites \cite{Lis2019}. In addition, the 67P/C-G measurements have revealed a Xe isotopic signature that differs significantly from anything else within the solar system, with an important depletion of heavy isotopes \cite{Marty2017, Avice2020ApJ}. This peculiar feature is shared with U-Xe, the primordial Xe component of the Earth's atmosphere (see \citet{Avice2020} for a review). Hence, it was concluded that the atmospheric Xe contains a 22\% contribution from comets, with the remainder having a chondritic origin \cite{Marty2017}. Yet, the xenon isotopic signature in the Earth's mantle appears to be chondritic \cite{Peron2018, Broadley2020}, even though measurement uncertainties do not preclude a small contribution from comets before the last giant impact. Either way, there is a dichotomy of cometary contributions to the Xe signature between the Earth's mantle and atmosphere, which naively indicates that most comets were brought to Earth after the Moon-forming impact and magma ocean solidification. This may at first glance seem contradictory with the \textit{Early instability} model, because the instability is known to increase the bombardment flux on Earth. However, in \cite{Joiret2023} we used numerical simulations to demonstrate how cometary impacts can occur tens of Myr after the instability due to the highly stochastic nature of the bombardment process. It is thus possible that enough cometary mass was brought to Earth after the Moon-forming impact so that the xenon constraint is not necessarily in conflict with an \textit{Early Instability}. Alternatively,  it is possible that the noble gases delivered by the comets were just not retained efficiently in the mantle and were sequestered in the atmosphere, but this hypothesis remains to be tested.

As far as the chondritic component in the Earth's atmosphere is concerned, the relative contributions of late accreted chondritic impactors or degassing of mantle-derived chondritic volatiles remains uncertain. \citet{Bekaert2020} calculated that a $\simeq$ 0.5 wt. \% of the Earth \textit{late veneer} contribution is not sufficient in supplying chondritic noble gases to the atmosphere, therefore requiring substantial contribution from mantle degassing. Consequently, a significant contribution from carbonaceous asteroids is expected before the last giant impact. 

\subsection{Aim of this study}
The large contrast in elemental and isotopic signatures between non-carbonaceous asteroids, carbonaceous asteroids and comets opens up the possibility of addressing the origin of Earth’s volatiles. In particular, geochemical constraints suggest that the dynamics of asteroids and comets might have changed with time and comets comprised less than $10^{-3}$ by mass of the late impacting population, but this calls for a theoretical validation \cite{Dauphas2002}. In this paper, we evaluate the relative contribution from carbonaceous asteroids and comets from a dynamical perspective, in the context of an \textit{Early Instability} model. We estimate the latter contributions before and after the Moon-forming impact and discuss the implications for the Earth's volatile budget. 

\section{Material and methods}
\subsection{N-body simulations}
High-resolution N-body simulations of the solar system are carried out using GENGA, a hybrid symplectic integrator, designed to integrate planet and planetesimal dynamics in the late stage of planet formation \cite{Grimm2014}. We start our sets of simulation at the time of gas disk dispersal ($t_{0}$) so we only consider gravitational forces. Massive bodies can either be treated as small or large bodies and follow different gravity modes. Large bodies are in full gravity mode, meaning forces between all pairs of particles are computed. Bodies with a smaller mass than a given threshold can either be treated as semi-active or test particles, depending on whether they interact with larger bodies only or do not interact with other bodies at all, respectively. In this study, we consider the terrestrial embryos to be large bodies, and planetesimals to be semi-active. Given that only one mass threshold can be fixed, carbonaceous asteroids (as massive as the planetesimals) and comets (less massive than planetesimals) are also treated as semi-active rather than test particles. Integrations last for 100 Myr after $t_{0}$.

\subsubsection{Initial conditions}
The initial conditions of our simulations are set to represent the early stages of solar system evolution leading to the constraints imposed by terrestrial planets, carbonaceous asteroids, giant planets, and comets. At the moment of gas disk dispersal ($t_{0}$), the giant planets are already formed. The inner solar system on the other hand comprises a bimodal population of planetesimals and Moon- to Mars-sized embryos, as predicted by analytical derivations and numerical modeling of rocky planet formation \cite{Greenberg1978, Wetherill1993, Kokubo1996}. The embryos were likely distributed into a ring centered between Venus and Earth's orbits. This initial configuration can be justified by several dynamical models including the Grand Tack \cite{Walsh2011} or the annulus model \cite{Drazkowska2017, Lichtenberg2021, Morby2022, Izidoro2022}, and by ALMA observations of protoplanetary disks \cite{Huang2018}. It also provides a potential solution to the small-Mars problem \cite{Raymond2009}. In our simulations, inner solar system embryos and planetesimals are uniformly distributed in a ring extending from 0.7 to 1.2 AU. We set a surface density profile that falls off as $\sum(r) = r^{-1}$. The initial eccentricities and initial inclinations follow a Rayleigh distributions with $\sigma_{e}$ = 0.005 and $\sigma_{i}$ = 0.0025 respectively, as considered in \citet{Nesvorny2021}.  We set the total mass of the ring to 2.5 $M_{\Earth}$, comprising 35 embryos of $\simeq$ 1/2 of Mars' mass and 2000 planetesimals of 1.25 $\times$ $10^{-4}$ $M_{\Earth}$ $\simeq$ $10^{-2}$ $M_{Moon}$ each. Indeed, \citet{Walsh2019} have shown that by the end of the gas disk lifetime (considering a gas disk density decreasing exponentially with a 2 Myr timescale) the population comprises nearly 90\% planetary embryos by mass around 1 AU. We assume a bulk density of 3 $g/cm^{3}$ for embryos and 2 $g/cm^{3}$ for planetesimals, which are typical measured values for rocky asteroids \cite{Carry2012}. These embryos and planetesimals are assumed to be entirely non-carbonaceous.

Regarding the carbonaceous asteroids, we consider 1600 bodies with the exact same mass as the locally-sourced planetesimals, amounting to a total of 0.2 $M_{\Earth}$ \cite{Raymond2022}. These asteroids are fairly realistic in size, as they are about the mass of Ceres. We assume a bulk density of 1.7 $g/cm^{3}$, the typical measured density for CI-chondrites \cite{Carry2012}. Carbonaceous asteroids are thought to be a byproduct of the giant planets formation \cite{Raymond2017}. As Jupiter and Saturn were growing, a fraction of the nearby volatile-rich planetesimals were scattered inwards, tending to reach high eccentricities \cite{Raymond2017}. Gas drag acts to damp those eccentricities but, depending on the strength drag – which is a combination of the planetesimal size and the timing of scattering – planetesimals may have ended up on low- or high-eccentricity orbits crossing the terrestrial planet zone \cite{Raymond2017}. In our simulations, we assume that most of the carbonaceous asteroids have high-eccentricity orbits at gas disk dispersal. They are uniformly distributed between 0.7 AU $\le$ $q = a(1-e)$ $\le$ 1.5 AU  \cite{OBrien2014} and $Q = a(1+e)$ $\le$ 5.5 AU (the semi-major axis of Jupiter at $t_{0}$ being 5.6 AU), even though they likely formed beyond the orbit of Jupiter \cite{Raymond2017}. The initial inclinations follow a Rayleigh distributions with $\sigma_{i}$ = 2.5 \cite{Raymond2017}.

We also consider an outer ring of cometary planetesimals uniformly distributed between 21 and 30 AU. The initial eccentricities and inclinations follow a Rayleigh distribution with $\sigma_{e}$ = 0.005 and $\sigma_{i}$ = 0.0025 respectively, and bulk density is set to 0.5 $g/cm^{3}$, which is the typical measured bulk density for cometary nuclei \cite{Richardson2007, Carry2012}. The mass of the primordial cometary reservoir is set to 25 $M_{\Earth}$, as suggested by models of the giant planet instability \cite{Nesvorny2018review} and accounting for losses by collisional evolution within the primordial reservoir \cite{Bottke2023}. There are 10000 comets with 2.5 $\times$ $10^{-3}$ $M_{\Earth}$ $\simeq 1.5 \times 10^{25}$ g $\simeq$ $M_{Pluto}$. It was shown that the primordial outer disk of comets should have had some hundreds to thousands Pluto-mass bodies, and they could have represented 10\% to 40\% of the total disk mass \cite{Nesvorny2016, Kaib2021}. Here, they represent 100\% of the total disk mass, because taking into account the effect of smaller comets would imply to increase the total number of particles in our simulation for a fixed outer disk mass. However, we are limited by computational costs associated with our simulations.

It should be noted that our choice of initial conditions is arbitrary, albeit inspired by previous studies \cite{OBrien2014, Raymond2017, Nesvorny2018review, Walsh2019, Nesvorny2021, Clement2021b, Raymond2022}. Quantitative results should thus be taken with caution, as there are large uncertainties on the initial state of our numerical simulations. However, our focus in this paper is the dynamics of the early evolution of the solar system, and whether this dynamics can match several geochemical constraints at the same time. 

\begin{figure}[h!]
\centering
  \includegraphics[width=11cm]{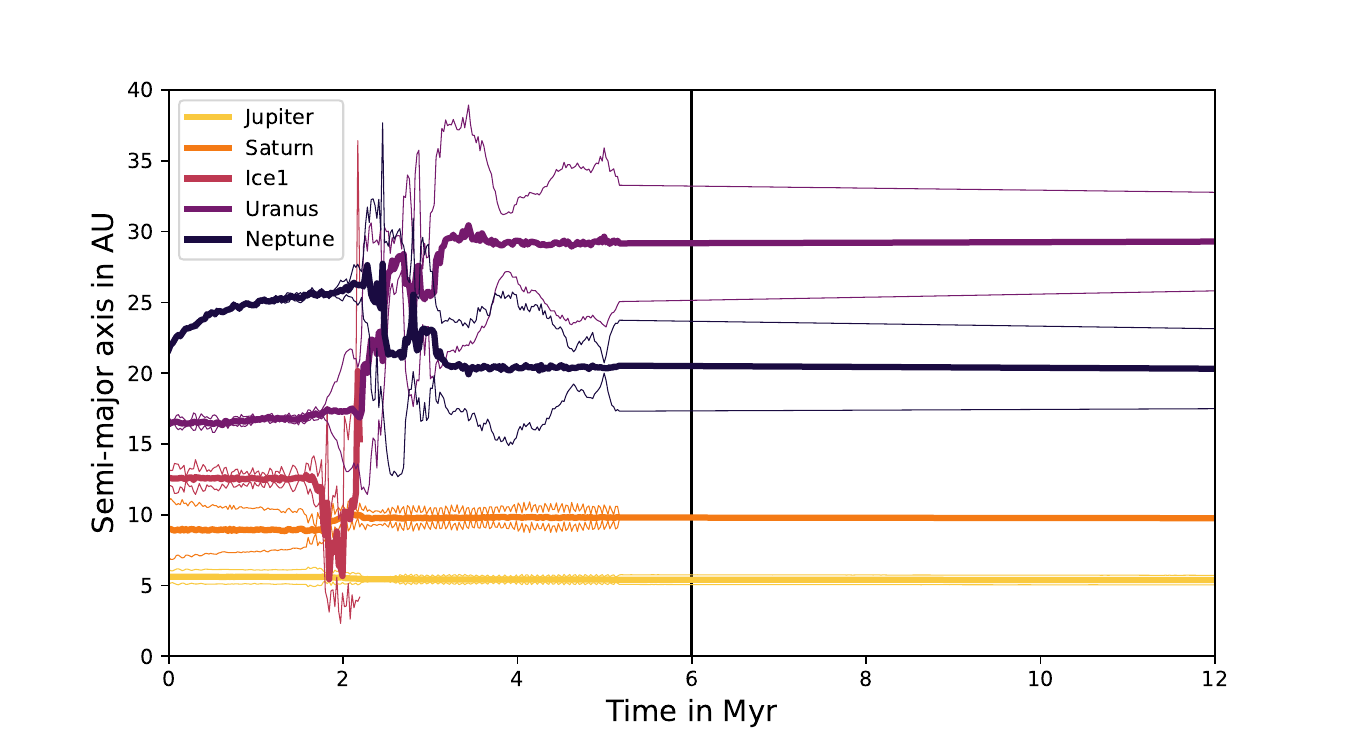}
  \caption{\small Early evolution of the giant planets in our simulations. The onset of the instability happens at $t_{inst}$ $\simeq$ 2 Myr. The extra ice giant is ejected right after $t_{inst}$ and the four remaining giants start stabilizing again. The black vertical line indicates the starting point of the second interpolation during which the giant planets evolve smoothly towards their current orbit. We only show the dynamical evolution of the first $\simeq$ 12 Myr for a visualization purpose. We refer the reader to \citet{Joiret2023} (Section 2.2.2) for more details about the instability.}
  \label{inst}
\end{figure}

We include the \textit{Early Instability} model \cite{Clement2018} by forcing the giant planets into a dynamical instability very early in their history. Most precisely, we force the giant planets to follow simulation outputs from \citet{Clement2021b}, where $t_{inst} \simeq$ 2 Myr (cf. Fig. \ref{inst}). In these simulation outputs, the giant planets start from an initial five-planet multi-resonant configuration with period ratios 2:1, 4:3, 3:2, 3:2 and large primordial eccentricities, which proved to be highly effective in producing a final system of planets with low eccentricities and inclinations. One planet is lost through the instability and the system is later stabilized by damping the excess orbital energy into the outer disk of planetesimals \cite{Nesvorny2011}. The mass of the additional ice giant is about half the mass of Uranus or Neptune, while the masses of the four other giant planets are set to their current values. In the separate integration conducted by \citet{Clement2021b}, the giant planets already interacted gravitationally with a 35 $M_{\Earth}$ outer disk, influencing their orbits. Forcing the giant planets in our simulations to evolve according to this separate integration, ensures that the outer solar system constraints will be satisfied so as to better focus on the most plausible evolution of the inner solar system, distribution of carbonaceous asteroids and outer disk of comets.

\subsubsection{Success criteria}
\label{success}
We measure the success of a simulation using the prescription presented in \citet{Joiret2023}, which is largely inspired by \citet{Nesvorny2021}. We apply the following success criteria:
\begin{enumerate}
    \item The outcome of the simulation must include two (and only two) planets with mass M $ \gtrsim \frac{M_{\Earth}}{2}$ between $\simeq$ 0.6 and 1.3 AU. These two planets are called the Venus and Earth analogs.
    \item Only one planet can be located between 1.3 and 2.0 AU, we call it a Mars analog. Its mass must be $\lesssim$ $\frac{M_{\Earth}}{3}$. However, the simulation can still be considered successful even in the absence of a Mars analog. 
    \item Only one planet can be located at a distance $<$ 0.6 AU, we call it a Mercury analog. Its mass must be $\lesssim$ $\frac{M_{\Earth}}{4}$. However, the simulation can still be considered successful even in the absence of a Mercury analog. 
\end{enumerate}

For every remaining successful simulations, we calculate the corresponding angular momentum deficit (AMD, or $S_{d}$) introduced by \citet{Laskar1997} and the radial mass concentration (RMC, or $S_{c}$) introduced by \citet{Chambers2001}, and compare it to the values of the solar system.

The aim of this present study is not to investigate the formation of the terrestrial planets in detail with a large suite of simulations, as it has already been done \cite{Clement2018, Clement2019, Nesvorny2021}. Since our objective is rather to study the bombardment at high resolution, we want to run a sufficient number of simulations that end up accurately matching the inner solar system constraints in a generalized manner.

\subsection{Collision rate calculations}
While the contribution of carbonaceous asteroids to a given Earth analog can directly be retrieved from collision statistics recorded by the integrator, the same does not hold true for comets. Indeed, cometary impacts in the inner solar system are much rarer. Hence, collision probabilities between comets and Earth analogs must be computed in a second step based on their respective orbits and radii. More specifically, at each timestep of a given simulation, we calculate collision rates between each of the 10000 comets and each of the constituent embryos/planetesimals that will form the Earth analog. For the sake of simplicity, we use the term 'Earth constituent embryo' even when we refer to Earth constituent planetesimals. 

The same can also be done between each of the 1600 carbonaceous asteroids and each of the Earth constituent embryos. The results can be compared to the true number of collisions recorded by a given simulation. This serves as an effective validation method of our collision rate algorithm.

The number of collisions between an Earth constituent embryo and a comet (or carbonaceous asteroid) occurring during a certain time interval $\Delta$t can be expressed as
\begin{equation}
    n_{coll} = P_{tot}(R+r)^{2} f_{grav} \Delta t
\end{equation}
where $P_{tot}$ is the collision probability per unit of time between two bodies when the sum of their respective radii equals 1 km. This so-called \textit{intrinsic} collision probability is obtained, together with the relative velocity between the two bodies, using an algorithm based on the {\"O}pik/Wetherill approach \cite{Wetherill1967, FarinellaDavis1992, Greenberg1982}, described in detail in Section 2.3 of \citet{Joiret2023}. R and r are the radii in km of the Earth analog constituent embryo and the comet (or carbonaceous asteroid) respectively, and $f_{grav}$ (dimensionless) is the gravitational focus of the Earth analog embryo given by 
\begin{equation}
    f_{grav} = 1 + \frac{v_{esc,e}^{2}}{U^{2}}
\end{equation}
with $v_{esc,e} = \sqrt{\frac{2GM_{e}}{R}}$ the escape velocity of the Earth analog embryo and U the relative velocity between the two bodies, also obtained from the {\"O}pik/Wetherill algorithm.

\section{Preliminary results}
\subsection{Diversity of outcomes}

\begin{table*}[t!]
\begin{center}
\begin{tabular}{lp{1.5cm}p{1.5cm}p{1.5cm}p{1.5cm}p{1.5cm}p{1.5cm}p{1.5cm}p{1.5cm}}
\hline \hline
    Simulation name & AMD & RMC & $M_{Earth}$ ($M_{\Earth}$) & $a_{Earth}$ (AU) & $M_{Venus}$ ($M_{\Earth}$)& $a_{Venus}$ (AU) & $M_{Mars}$ ($M_{\Earth}$)& $a_{Mars}$ (AU)\\
 \textbf{solar system} & \textbf{0.0018} & \textbf{89.9} & \textbf{1} & \textbf{1} & \textbf{0.815} & \textbf{0.723} & \textbf{0.107} & \textbf{1.524}\\ \hline
 CL1 & 0.0011 & 101.3 & 0.962 & 1.120 & 1.284 & 0.705 & / & / \\ 
 CL2 & 0.0037 & 79.7 & 1.186 & 1.027 & 0.645 & 0.637 & 0.067 & 1.739 \\ 
 CL6 & 0.0014 & 76.8 & 1.309 & 1.126 & 0.932 & 0.661 & / & / \\ 
 CL13 & 0.0037 & 76.9 & 0.904 & 1.036 & 1.150 & 0.649 & 0.068 & 1.653 \\
 CL14 & 0.0042 & 57.5 & 0.964 & 1.259 & 1.352 & 0.680 & / & / \\
 CL16 & 0.0022 & 54.3 & 0.966 & 1.336 & 1.3328 & 0.709 & / & / \\
 CL17 & 0.0049 & 60.4 & 1.176 & 1.141 & 1.074 & 0.631 & / & / \\ 
 CL18 & 0.0116 & 84.4 & 0.687 & 1.110 & 1.418 & 0.687 & 0.069 & 1.690 \\ 
 CL19 & 0.0027 & 72.6 & 1.253 & 0.996 & 0.865 & 0.634 & 0.140 & 1.632 \\
 CL20 & 0.0030 & 44.6 & 1.042 & 1.002 & 1.004 & 0.614 & 0.271 & 1.754 \\ \hline \hline
\end{tabular}
\caption{\small Summary of all successful simulations (i.e. meeting the criteria established in section \ref{success}). AMD and RMC stand for angular momentum deficit and radial mass concentration respectively. The AMD, RMC, masses and semi-major axes of the terrestrial planet analogs are taken at the end of each simulation (i.e. at t = 100 Myr). The values specific to the solar system are displayed at the top for comparison.}
\end{center}
\label{table}
\end{table*}

We perform 20 simulations and half of them end up being successful according to the list of criteria in section \ref{success}. A summary of these successful outcomes can be found in Table \ref{table}. Each simulation includes both a Venus and a Earth analog, 5 of them include a Mars analog and none of them include a Mercury analog. 

While the mass of the Earth analogs oscillates around the expected value of 1 $M_{\Earth}$, the Venus analogs grow too large in most simulations. Additionally, the semi-major axis of the Venus analogs ends up too close to the Sun, whereas the Earth and Mars analogs are too far. This leads to lower RMC values (cf. Table \ref{table}). This issue is common in those types of simulations, as noted by other studies \cite{Nesvorny2021}. This might also be related to the fact no Mercury analogs were formed either. A different structure in the inner disk is needed to match Venus and Mercury well \cite{Clement2023, Lykawka2023}. One potential solution to this could be to start with a tighter inner ring of planetesimals in our initial conditions. For example, terrestrial embryos and planetesimals could be uniformly distributed in a ring extending from 0.8 to 1.1 AU at $t_{0}$, rather than from 0.7 to 1.2 AU. Starting with embryos that are larger could also help increase the RMC, but this might in turn result in unrealistically short growth timescales of the terrestrial planets. 

Table \ref{table2} shows that for all simulations except CL1, CL13, CL18 and CL20, the last giant impact happens in the [30-100] Myr range after gas disk dispersal. It is important to note that the time of the Moon-forming impact is usually constrained considering that $t_{0}$ is the condensation of the first solids in the solar system \cite{Kleine} - which predates the time of gas disk dispersal. The gas disk lifetime, that is a few Myr \cite{Haisch2001, Wang2017}, should thus be added to our values of the Moon-forming impact time for comparison with the literature. The Moon-forming impactor mass is often constrained to $\simeq$ 0.1 - 0.11 $M_{\Earth}$ \cite{Canup2001}, even though a wider range of initial conditions is possible if we consider that the Earth was spinning faster at the end of its accretion \cite{Cuk2012, Canup2021}. In 6 out of 10 simulations, we find a mass comparable to that of Mars for the last giant impactor. Simulations CL2, CL14, CL17 and CL19 are particularly successful at replicating most solar system constraints, including those of the Moon-forming impact.

In table \ref{table2}, we also show the mean-life of accretion, $\tau$, corresponding to the time taken to accrete 63\% of the Earth’s mass \cite{Jacobsen2005}. Most of the Earth analogs in our simulations have a typical $\tau \simeq$ 10 Myr. Using the exponential growth model \cite{Wetherill1986} and assuming that core formation was an equilibrium process - i.e. if the cores of impactors completely emulsified as small metal droplets in the terrestrial magma ocean -, these accretion timings would be consistent with Hf–W data \cite{Jacobsen2005, Kleine}. However, it is more likely that only 40\% of the cores of accreted objects equilibrated with Earth’s mantle before entering Earth’s core \cite{Rudge}, in which case $\tau$ should be closer to $\simeq$ 40 Myr \cite{Kleine2011}. The growth's timing of our Earth's analogs is thus consistently too short, except in simulation CL16 where it is too long. This issue is not critical in the present study, as the timing of Earth’s growth do not influence the influx of carbonaceous asteroids or comets.

\begin{figure}[h!]
\centering
  \includegraphics[width=11cm]{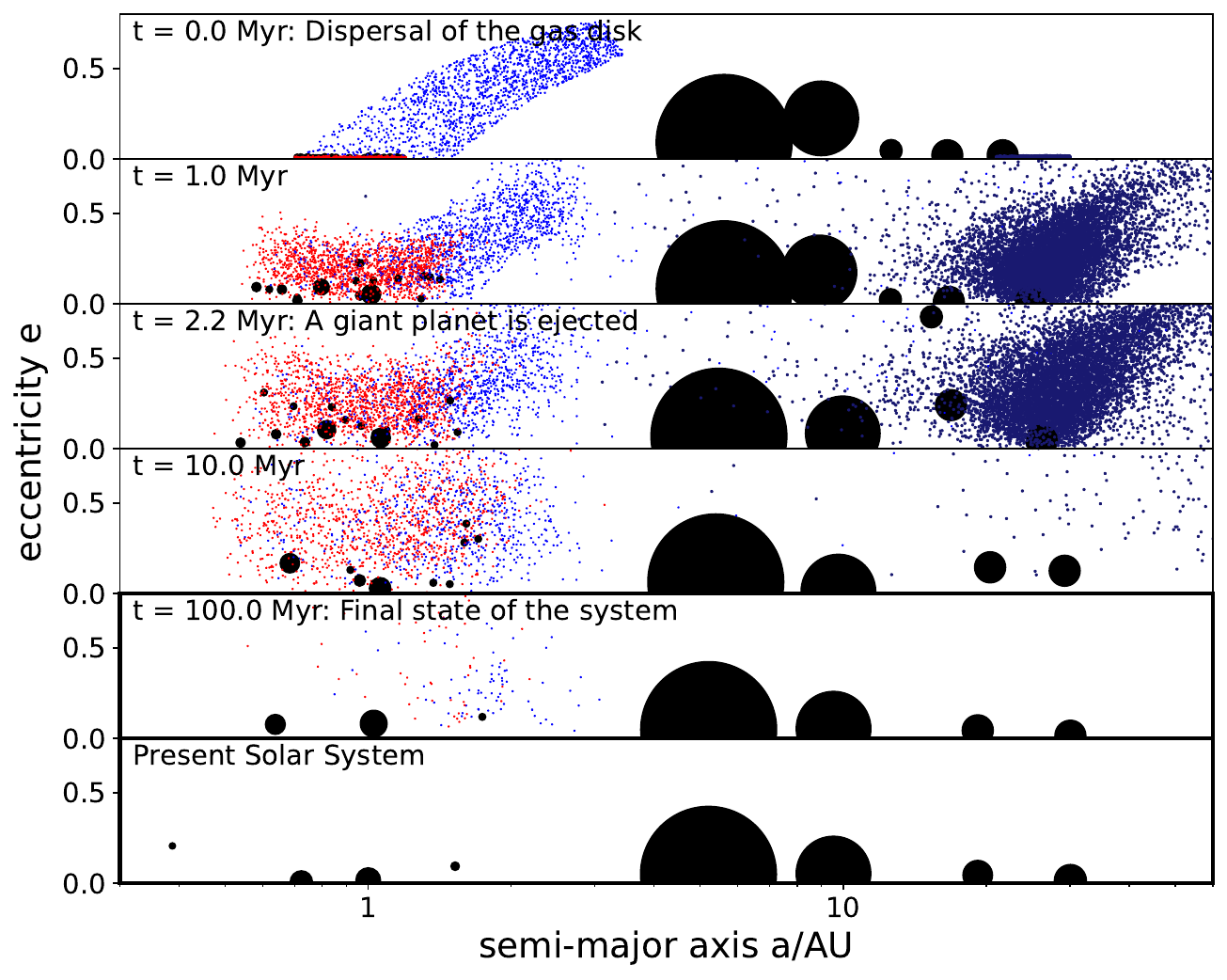}
  \caption{\small Simulation CL2. The symbol sizes are proportional to the mass of the bodies. Terrestrial embryos/planets and giant planets are in black, locally-sourced planetesimals are in red, carbonaceous planetesimals are in blue and comets are in dark blue.  For the giant planets a fixed size was chosen so an not to fill most of the panels.}
  \label{evol}
\end{figure}

Fig. \ref{evol} illustrates the temporal evolution of one successful simulation (CL2). We compare its final state to the present solar system (lower panel). The upper panel shows the start of the simulation, at gas disk dispersal, with an inner ring of dry planetesimals, a distribution of carbonaceous asteroids that came from the outer parts of the solar system and were deposited in the inner region while Jupiter and Saturn were growing \cite{Raymond2017}, 5 fully-grown giant planets, and an outer ring of comets. During the giant planet instability, most of the comets are ejected outwards, but some of them are ejected inwards and reach the inner solar system. The distribution of carbonaceous asteroids is also destabilized and many of these volatile-rich asteroids collide with the terrestrial planets throughout the simulation. At t = 2.2 Myr (third panel), one of the giant planets is ejected out of the solar system. By the end of the instability, the outer ring of comets is largely depleted. In this simulation, the Moon-forming impact happens at t = 49.8 Myr when a Mars-sized body impacts the Earth analog. The integration ends at t = 100 Myr (second lower panel). At that point, the inner solar system is comprised of 3 rocky planets with masses and orbits comparable to those of the present Venus, Earth and Mars respectively.

\subsection{Validation of the collision rate code}
\label{validation}
\begin{figure}[h!]
\centering
  \includegraphics[width=16cm]{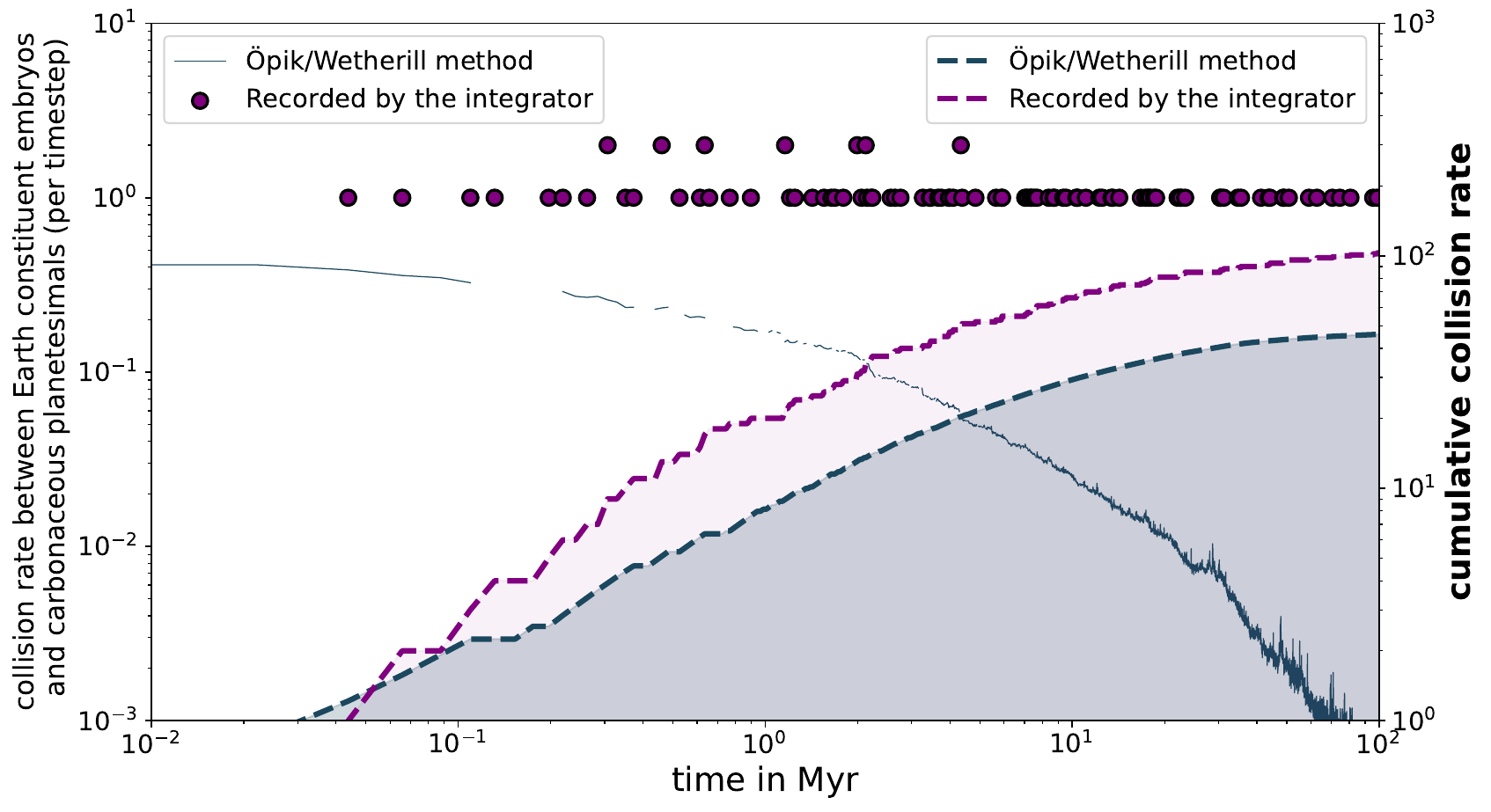}
  \caption{\small Comparison between the collision rate of the carbonaceous planetesimals obtained from the {\"O}pik/Wetherill method and recorded by the integrator in simulation CL1.}
  \label{validationfig}
\end{figure}

The contribution of carbonaceous planetesimals to a given Earth analog can either be calculated from collision statistics recorded by the integrator or using the {\"O}pik/Wetherill algorithm. We report the collision rates obtained from both methods in simulation CL1 (see Fig. \ref{validationfig}). It can be seen that the respective cumulative collision rates - corresponding to the number of collisions by the end of the simulation - are very close. While the {\"O}pik/Wetherill approach represents a satisfactory approximation of the statistics recorded by the integrator, it might underestimate by a factor of 2 the total number of collisions. In this study, this may have a small significance on the collision rates of comets, as they can solely be obtained from the algorithm. In addition, the {\"O}pik/Wetherill algorithm may exhibit larger errors when applied to comets, as their highly elongated orbits may not align well with certain assumptions of the algorithm. Collision rates of carbonaceous planetesimals on the other hand are taken from the integrator for the rest of the paper. It is also important to note that the main conclusions of this study rely on orders of magnitude and remain unaffected by this factor 2 difference. 

\section{Contribution from carbonaceous asteroids}
\label{CC_cont}
\begin{figure}[h!]
\centering
  \includegraphics[width=14cm]{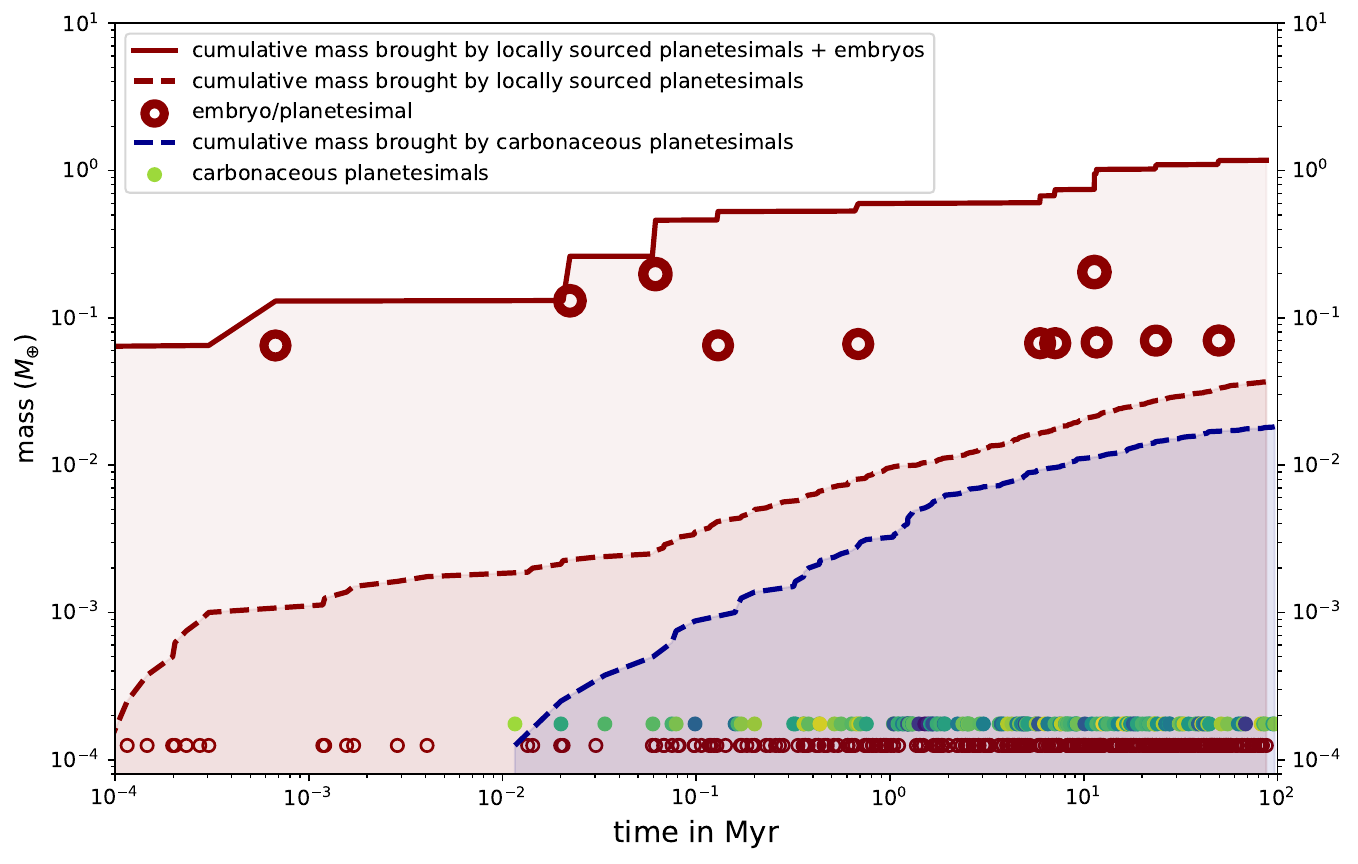}
  \caption{\small Mass brought by locally-sourced and carbonaceous material to the Earth analog in simulation CL2, as recorded by the integrator. Carbonaceous planetesimals have the same mass as locally-sourced planetesimals (1.25 $\times$ $10^{-4}$ $M_{\Earth}$), but are displayed above them on the plot to avoid overlapping. Carbonaceous planetesimals are coloured according to their semi-major axis at gas disk dispersal. This latter parameter is not representative of their volatile content as they likely formed further out in the solar system (between 4 and 9 AU) and were ejected into the inner solar system as Jupiter and Saturn were growing, before gas disk dispersal \cite{Raymond2017}. Although there is no trend between semi-major axis of carbonaceous planetesimals at gas disk dispersal and collision time with the Earth analog, locally-sourced (non-carbonaceous) planetesimals start colliding with the Earth analogs before carbonaceous planetesimals do. In this simulation, there is a 1.5 wt.\% contribution of carbonaceous material to the Earth analog.}
  \label{as_truecoll}
\end{figure}

We estimate the relative contributions from non-carbonaceous (locally-sourced) and carbonaceous material to Earth analogs. In total, carbonaceous planetesimals contribute between 1 and 2 wt.\% of the fully-grown Earth analogs (cf. Table \ref{table2} and Fig. \ref{as_truecoll}). This is in good agreement with the finding that the Earth's water and highly volatile budget requires the accretion of $\simeq$ 2 ($\pm$1) wt.\% of carbonaceous (CI or CM-like) material \cite{Marty2012}. More recent works constrained this carbonaceous contribution to about 4\% \cite{Burkhardt2021} or 6\% \cite{Savage2022} assuming CI chondrite-like material, and up to 10\% assuming CV chondrite-like material \cite{Kleine2023}. 
Considering the initial mass of carbonaceous asteroids in our simulations is a free parameter, we could simply increase it by a factor of 3 or more in order to match the latter constraints. However, in that case, and assuming the dynamics did not change, there would be 3 times more carbonaceous material hitting Earth at all times. This would thus be inconsistent with the finding that the late accretion was mostly non-carbonaceous \cite{Worsham2021} (cf. Fig. \ref{carb_LV}). From our simulations, it appears impossible to match both the total carbonaceous fraction and the non-carbonaceous-dominated late accretion at the same time, except if a significant contribution from carbonaceous material happened before gas disk dispersal. As \citet{Raymond2017} have shown, there were likely several episodes of inward scattering of carbonaceous material during the giant planets' growth and migration - so before gas disk dispersal. It is plausible that the earliest one, i.e. associated with Jupiter's growth, could have scattered inward some carbonaceous material with low volatile content because it may have formed early and have been dried up by Al-26 heating \cite{Monteux2018}. A major portion of carbonaceous material would thus have been delivered to Earth early, before gas disk dispersal. A single small carbonaceous embryo could have been sufficient in order to match a 4-10\% carbonaceous contribution to Earth. If only a small number of carbonaceous embryos delivered volatiles, including water, to Earth, then Mars could have avoided a collision and have remained much drier than Earth. We leave the detailed exploration of carbonaceous delivery to the inner disk for future work.

The early delivery hypothesis was notably invoked to explain the deep mantle Kr isotopic signature \cite{Peron2021} or the Ru and Mo isotopic compositions of lunar impactites \cite{Worsham2021}, even though it is more commonly accepted that water–bearing bodies were accreted preferentially late during Earth’s growth \cite{OBrien2014, Rubie2015, Liu2023, Dauphas2024}. Fig. \ref{as_truecoll} shows results from simulation CL2, where the Earth analog ends up with a mass of 1.186 $M_{\Earth}$ among which 1.825 $\times$ $10^{-2}$ $M_{\Earth}$ is carbonaceous material ($\simeq$ 1.5 wt.\% of the Earth analog).

\begin{figure}[h!]
\centering
  \includegraphics[width=14cm]{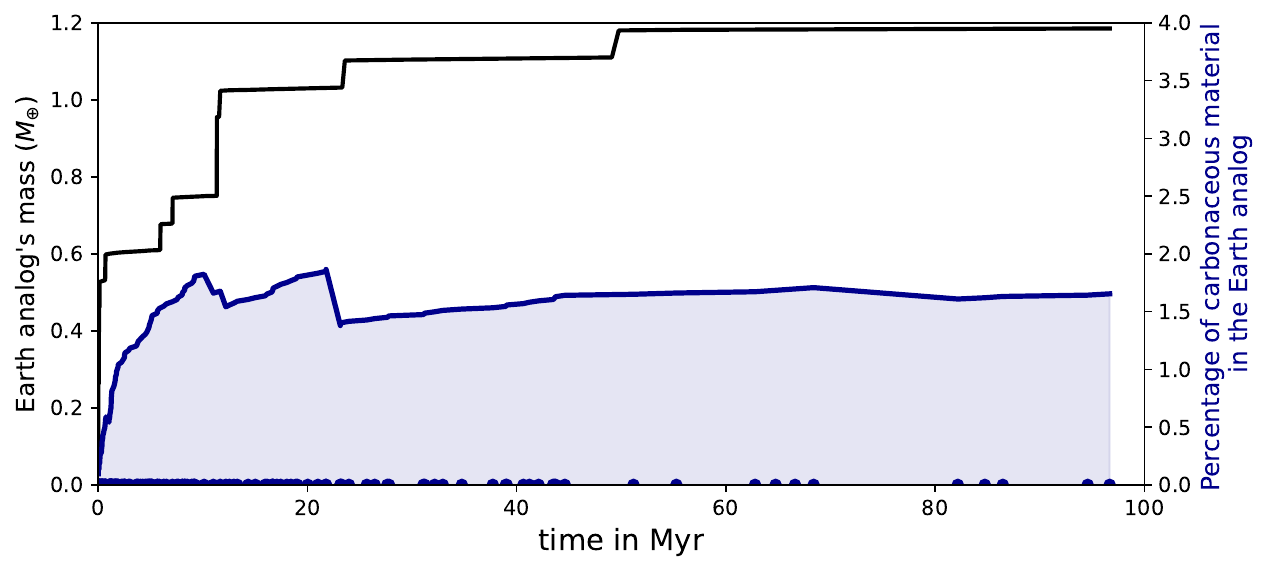}
  \caption{\small Percentage of carbonaceous material relative to the Earth analog's mass as a function of time in simulation CL2. }
  \label{carb}
\end{figure}

Fig. \ref{carb} illustrates the temporal evolution of the carbonaceous fraction in the Earth analog of simulation CL2. It appears that this carbonaceous proportion is quite regular and it cannot be concluded from our simulations that the carbonaceous contribution was added preferentially late in the accretion process, as suggested by \citet{Dauphas2024}. We observe a similar trend in all our simulations. Nevertheless, collisions from carbonaceous asteroids start when the Earth analogs reach $\simeq$ 10 - 30\% of their final mass, in all our simulations. So there is always at least one embryo-embryo impact before any carbonaceous impacts, after gas disk dispersal.

\begin{figure}[h!]
\centering
  \includegraphics[width=16cm]{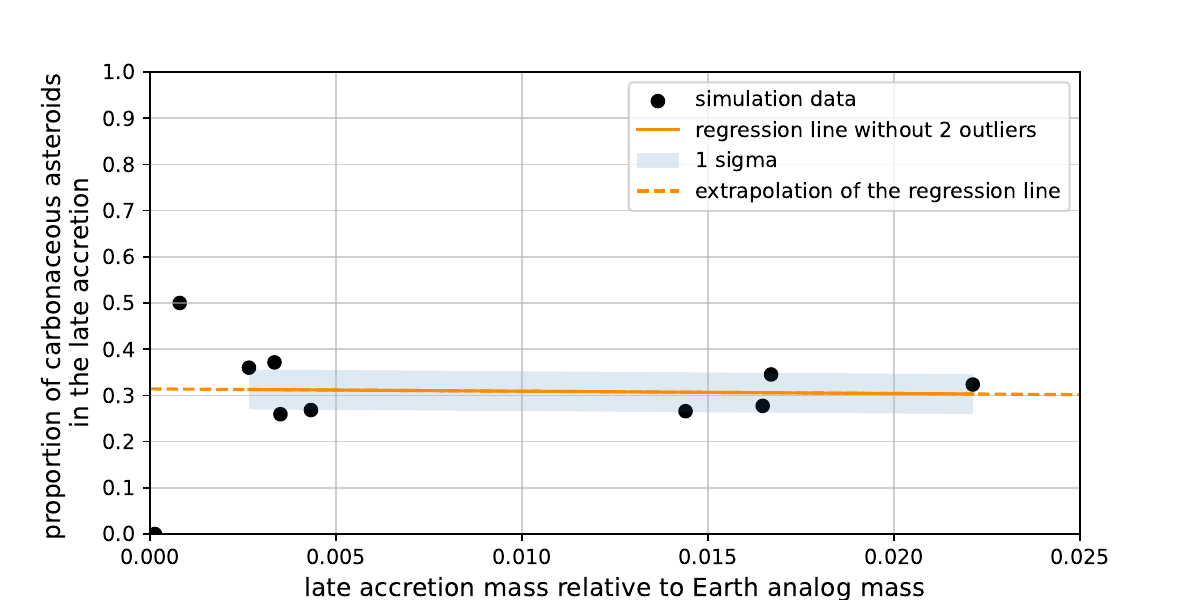}
  \caption{\small Contribution from carbonaceous asteroids after the last giant impact. We removed two outliers corresponding to simulations where the late accreted mass was too small considering our asteroids' mass resolution. We find a best estimate (with 1-$\sigma$ uncertainty) of 31.4\% $\pm$ 4.3\%.}
  \label{carb_LV}
\end{figure}

We find that no matter the time of the Moon-forming impact, about 30\% of the late accretion (precisely 31.4\% $\pm$ 4.3\%) is comprised of carbonaceous chondrites (Fig. \ref{carb_LV}). This result is consistent with the finding that about 30\% of the bulk silicate Earth’s Mo was delivered by carbonaceous material \cite{Burkhardt2021}.  As a moderately siderophile element, Mo only records the later stages of Earth's accretion \cite{Dauphas2017}. Similarly, Zn, a slightly siderophile element \cite{Siebert2011, Mahan2017}, is also expected to have been delivered late as it would otherwise have partitioned into the core during differentiation \cite{Mahan2018}. About a third of Earth’s Zn was delivered by a carbonaceous-like component \cite{Sossi2018, Savage2022}. On the other hand, Ru and Mo isotopic signatures of lunar impactites reveal that only up to $\simeq$ 20 to 25\%, and potentially much less, of carbonaceous material account for the late accretion mass on Earth \cite{Worsham2021}. This holds true if the impactites are representative of late accretion in general.

Considering that the Earth accreted 0.5\% of its final mass after the Moon-forming impact \cite{Walker2009, Jacobson, Morby2015}, this result would imply that a carbonaceous mass of 0.3 $\times$ 0.5\% = 0.15 wt.\% of the Earth was delivered during the late accretion. If the total mass contribution of carbonaceous asteroids to Earth is about 1.5 wt.\% (cf. Table \ref{table2}), 90\% of the carbonaceous chondrites mass was brought before the Moon-forming impact. If instead the total carbonaceous contribution is close to 4 wt.\% \cite{Burkhardt2021} or 6 wt.\% \cite{Savage2022}, then more than 96\% of the carbonaceous chrondrites were brought before the Moon-forming impact.

\subsection{Time of the Moon-forming impact}

\begin{figure}[h!]
  \centering
  \includegraphics[width=15cm]{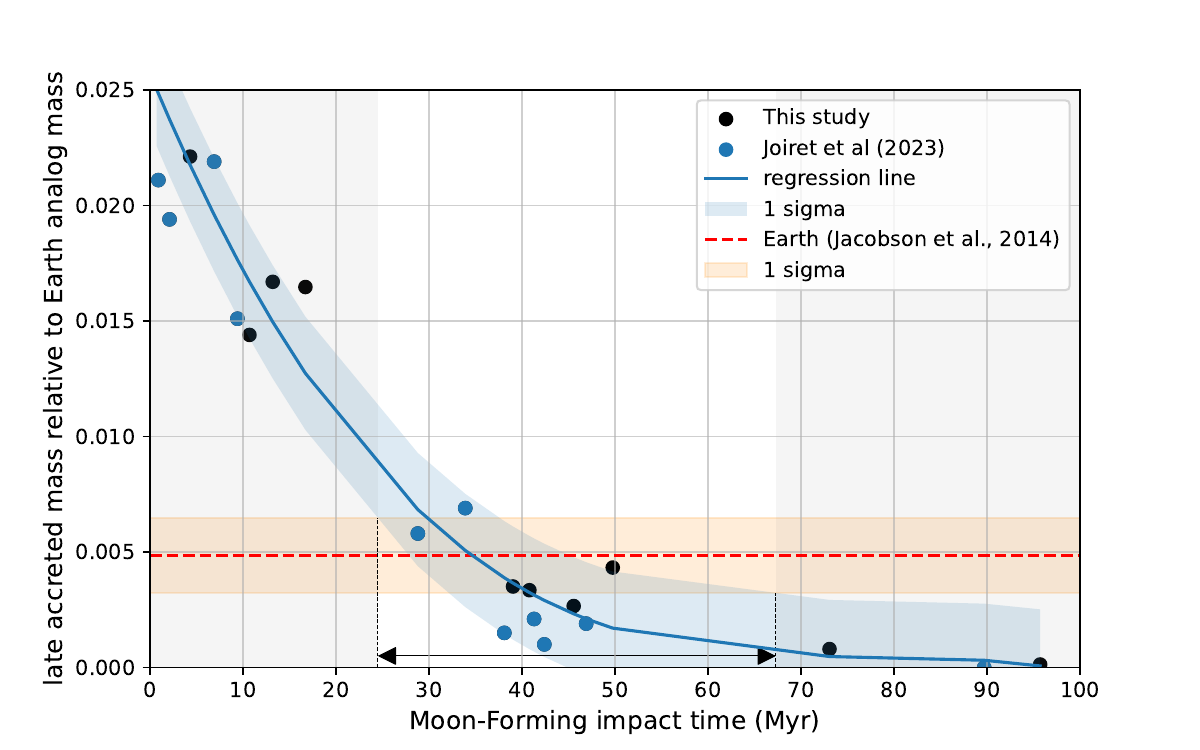}
  \caption{\small The late accreted mass relative to the Earth analog mass as a function of the time of the last giant impact. The red dashed line is the best estimate (with 1-$\sigma$ uncertainty) of the \textit{late veneer} mass inferred from the HSE abundances in the mantle: 4.8 $\pm$ 1.6 $\times$ $10^{-3}$ $M_{\Earth}$ \cite{Jacobson}. The Moon-forming impact time is calculated considering that $t_{0}$ is the moment of gas disk dispersal. If we considered $t_{0}$ as the condensation of the first solids in the solar system, all the simulation data - and the corresponding regression line -  would be translated a few million years to the right \cite{Haisch2001, Wang2017}.}
  \label{MFI_time}
\end{figure}

The sooner the last giant impact, the larger the late accreted mass \cite{Jacobson} - which is by definition the mass accreted on Earth after the last giant impact. The inverse correlation between the time of the Moon forming impact, i.e. the last giant impact, and the late accreted mass is shown in Fig \ref{MFI_time}. Our simulations data (including those of \citet{Joiret2023}) preclude with 1-$\sigma$ uncertainty a Moon-forming impact happening before 24 Myr or after 67 Myr after gas disk dispersal. The Moon-forming impact time range is often referred to considering that $t_{0}$ is the condensation of the first solids in the solar system. Here we consider $t_{0}$ as the time of gas disk dispersal. If we consider a gas disk lifetime of 4 Myr for example \cite{Haisch2001, Wang2017}, our best estimate of the Moon-forming impact is [28-71] Myr within 1-$\sigma$ uncertainty. This lower bound is reasonably consistent with the Hf‐W systematics that constrain the Moon-forming impact to after 30 Myr \cite{Nimmo2015, Thiemens2019}. However, our results should be interpreted with caution as they depend on at least two uncertain variables. 

First, the amount of mass that the Earth receives during late accretion is related to the size frequency distribution of the asteroids in the simulations. If the asteroid distribution is limited to Ceres-size objects or smaller, the total amount of late accreted mass by the Earth is less than $10^{-3}$ $M_{\Earth}$ \cite{Marchi2014}. This argument is used to support that the late accretion was dominated by a stochastic regime, in which a few very large projectiles determined the Earth's HSEs budget \cite{Bottke2010, Raymond2013, Marchi2014, Brasser2016}. It is also argued that $\ge$1000 km asteroids are essential to explain the HSE abundances in the martian mantle \cite{Marchi2020}. In this study, the size of asteroids (non-carbonaceous and carbonaceous) is about 1000 km across. Yet we find a direct relation between late accretion mass and time of the Moon-forming impact, as already shown by \citet{Jacobson}. 

Second, the expected \textit{late veneer} mass of 4.8 $\pm$ 1.6 $\times$ $10^{-3}$ $M_{\Earth}$ was inferred based on HSE abundances in the mantle \cite{Jacobson}. However, this mass could slightly differ from the late accretion mass as the process of Earth's differentiation probably happened episodically and stochastically \cite{Marchi2018}, and some HSEs from the Earth's mantle could have been accreted prior to the Moon-forming event \cite{Morby2015}. In fact, the degree of equilibration between metal and silicate after each impact depends on many factors including the size and velocity of the impactor, and whether it is differentiated or not \cite{Deguen2011}. It is demonstrated that the \textit{late veneer} mass - defined through the abundance of HSEs in the silicate Earth - and the late accretion mass - the actual mass that was accreted to Earth after the Moon-forming impact - are the same within a factor of 2 \cite{Morby2015}. More precisely, the Earth should not have accreted more than 1 wt.\% after the Moon-forming event and it is possible that a fraction of the HSEs, perhaps as much as $\simeq$ 50\% (i.e. $\simeq$ 0.25 wt.\% of the Earth), predates the Moon-forming event \cite{Morby2015}. If indeed the late accretion mass represented half of the \textit{late veneer} mass, the upper bound of the Moon-forming impact time range would thus be greater than 100 Myr. Conversely, if the late accretion mass amounts to twice the late veneer mass, this upper bound would be less than 40 Myr. 

\subsection{Contribution from carbonaceous asteroids to Mars}
We find that carbonaceous asteroids contribute $\simeq$ 1 - 2 wt.\% of the fully grown Mars analogs. This result is fairly consistent with the Zn isotope data that limits the contribution of carbonaceous material (CI-like) to Mars to 4 wt.\% at most, with a best fit fraction of 0.5 wt.\% \cite{Kleine2023}. 

Considering that we find similar proportions for the total carbonaceous contribution on Earth and Mars, but Mars is expected to have accreted less carbonaceous material than Earth \cite{Kleine2023}, we support the idea that Mars was less able to retain volatiles upon impacts. This could be related to its smaller planetary mass \cite{Young2019}, and potentially much smaller atmosphere, which would have been less efficient at decelerating and ablating impactors.

\section{Contribution from comets}
\begin{figure}[h!]
  \centering
  \includegraphics[width=12cm]{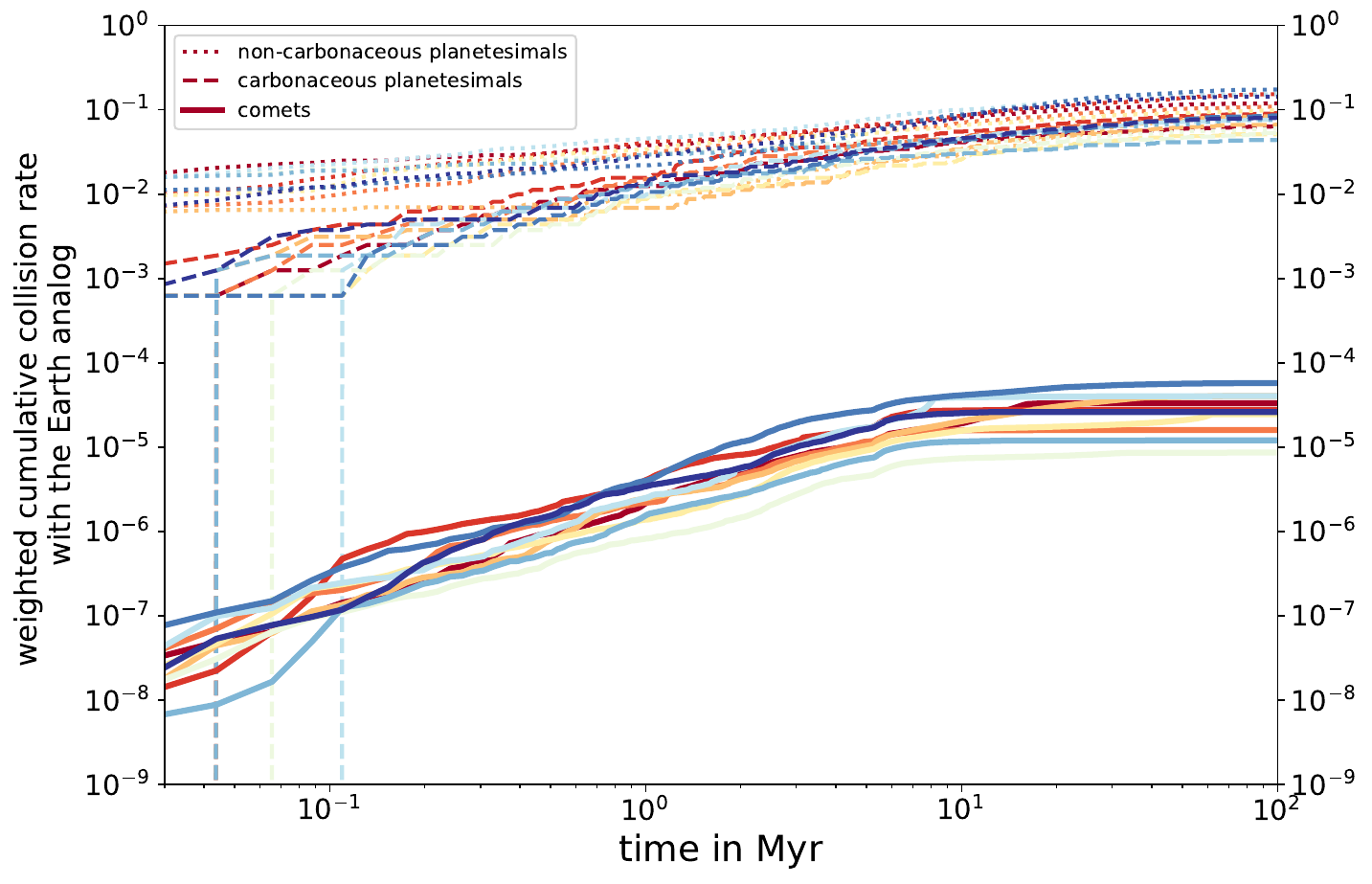}
  \caption{\small Comparison of the weighted average cumulative collision rate of a non-carbonaceous (locally-sourced) planetesimal (dotted lines), a carbonaceous planetesimal (dashed lines), and a comet (solid lines) respectively, with an Earth analog. These lines are obtained by dividing the cumulative collision rate of NC/CC/comets in each simulation with the number of NC/CC/comets respectively in these simulations. Each color represents a different simulation.}
  \label{comp_ascom}
\end{figure}

Fig. \ref{comp_ascom} shows the cumulative collision rates of non-carbonaceous (locally-sourced) planetesimals, carbonaceous planetesimals and comets with the Earth constituent embryos. The cumulative collision rates - i.e. the number of collisions occurring in a 100 Myr time period - are weighted by the number of non-carbonaceous planetesimals (2000), carbonaceous planetesimals (1600) and comets (10000) respectively in our simulations. A cumulative collision rate of $\simeq$ 0.1 for the carbonaceous planetesimals after 100 Myr means that if we start with $10^{5}$ carbonaceous planetesimals at $t_{0}$, there will be about 0.1 $\times$ $10^{5}$ = $10^{4}$ collisions between carbonaceous planetesimals and the Earth analog by the end of the simulation. Similarly, a cumulative collision rate of $\simeq$ $10^{-5}$ for the comets after 100 Myr, means that if we start with $10^{5}$ comets at $t_{0}$, there will be about $10^{-5}$ $\times$ $10^{5}$ = 1 collision between a comet and the Earth analog by the end of the simulation. 

The cumulative collision rates of carbonaceous planetesimals with Earth are $\lesssim$ 1 order of magnitude smaller than those of non-carbonaceous planetesimals - not embryos - before the first Myr and this difference slightly decreases with time. 

When we compare the cumulative collision rate of planetesimals with the Earth analog and comets with the Earth analog, there is a much larger gap. A collision with a comet is more than 4 orders of magnitude less likely than a collision with a carbonaceous asteroid, and it is 4 - 5 orders of magnitude less likely than a collision with a non-carbonaceous asteroid. These numbers are calculated for one comet or one asteroid only, as the cumulative collision rates are weighted by the number of comets (10000) and asteroids (1600 for the carbonaceous, and 2000 for the non-carbonaceous) respectively in our simulations. That means the comet/asteroid collision rate ratio would increase if there were initially many more comets than asteroids, and decrease otherwise. 

$^{132}Xe$ is approximately 3 orders of magnitude more abundant in comets than in carbonaceous asteroids, and 3 - 4 orders of magnitude more abundant in comets than in non-carbonaceous asteroids \cite{Bekaert2020}. However, the relative cumulative collision rates may have canceled out the higher abundance of Xe in comets, or have allowed up to 10\% of cometary Xe at most in the Earth's mantle. From these considerations, it is easy to grasp how the xenon isotopic signature in the Earth’s mantle turned out to be chondritic \cite{Peron2018, Broadley2020}. However, the xenon isotopic dichotomy between the Earth's mantle and atmosphere remains to be understood. 

\begin{figure}[h!]
  \centering
  \includegraphics[width=12cm]{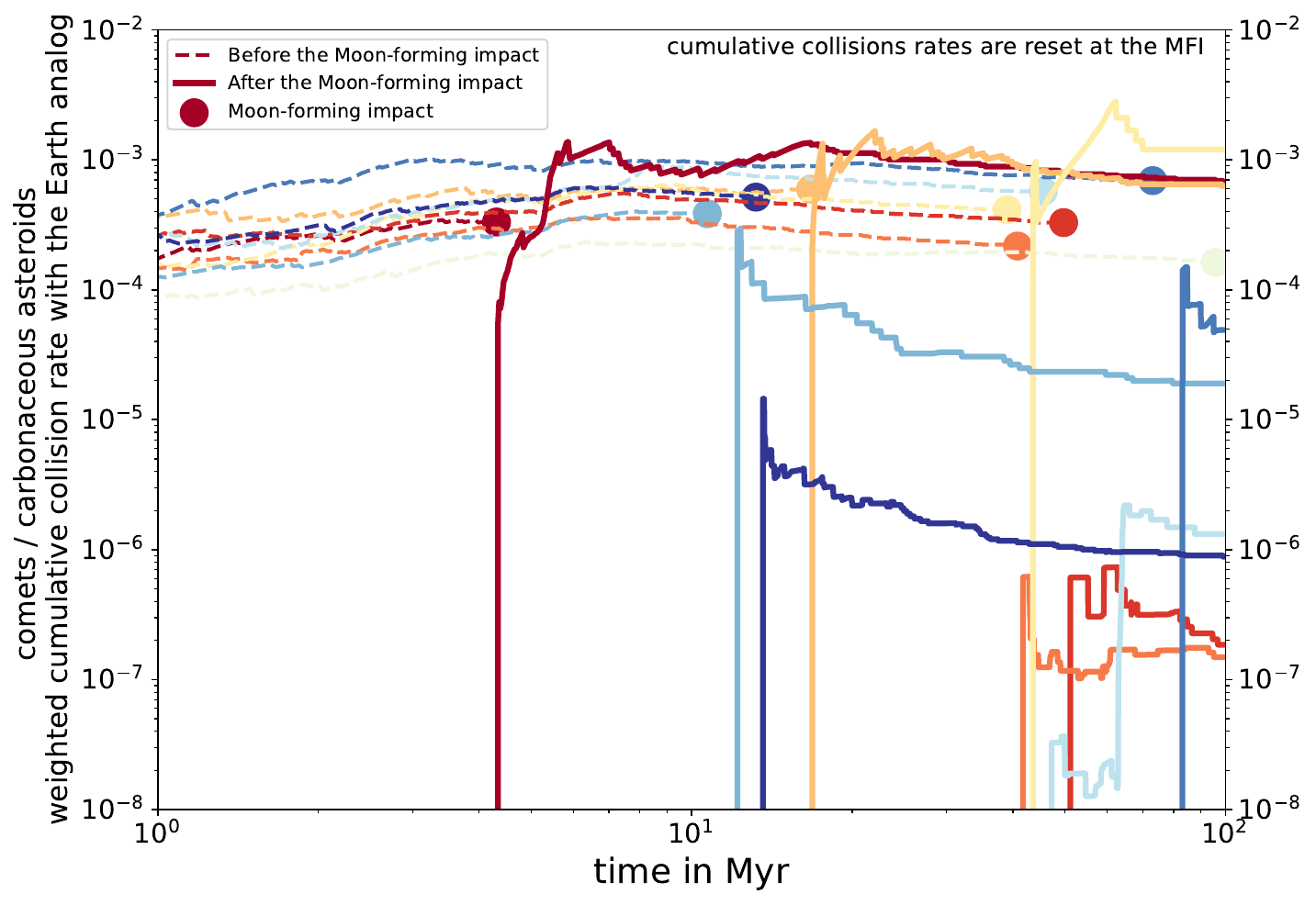}
  \caption{\small Ratio of the cumulative collision rate of comets with the Earth analog to the cumulative collision rate of carbonaceous asteroids with the Earth analog. The cumulative collision rates are reset at the time of the Moon-forming impact; and the ratio is set to 0 when the cumulative collision rates of carbonaceous asteroids (denominator) equals 0. So the dashed line shows the number of collisions with comets relative to the number of collisions with carbonaceous asteroids (weighted by their respective numbers in our simulations) on Earth since gas disk dispersal until the Moon-forming impact; and the solid line shows the same relative number since the Moon-forming impact until the end of the simulation. While cumulative collision rates of comets with Earth are $\simeq$ 4 orders of magnitude smaller than those of carbonaceous asteroids \textit{before} the Moon-forming impact, they are between 3 and 7 orders of magnitude smaller \textit{after} the Moon-forming impact. In 30\% of our simulations, there is about 1 comet every $10^{3}$ carbonaceous asteroids colliding with the Earth after the Moon-forming event, which is consistent with the total cometary mass expected during late accretion.}
  \label{comp_ascom_MFI}
\end{figure}

Fig. \ref{comp_ascom_MFI} shows the ratio of the weighted cumulative collision rate of comets with the Earth analog to the weighted cumulative collision rate of carbonaceous asteroids with the Earth analog (basically, the ratio of the solid line to the dashed line in Fig. \ref{comp_ascom} before and after the Moon-forming event). Before the Moon-forming impact (dashed line in Fig. \ref{comp_ascom_MFI}), this ratio is roughly constant and is generally close to $10^{-4}$. In other words, a collision with a comet is $\simeq$ 4 orders of magnitude less likely than a collision with a carbonaceous asteroid. After the Moon-forming impact (solid line in Fig. \ref{comp_ascom_MFI}), the ratio varies between $10^{-3}$ and $10^{-7}$. This large dispersion of values is due to the fact that the cumulative collision rate of asteroids with Earth slightly decreases after the Moon-forming impact, while the one of comets either decreases or remains steady. This is a product of the stochastic component of the cometary bombardment \cite{Joiret2023}.
Thus, in 7 simulations out of 10, the ratio of the cumulative collision rates of comets relative to asteroids decreases after the Moon-forming impact. This scenario would lead to a chondritic Xe signature in the Earth's atmosphere, which is not consistent with the composition of our atmosphere. In 3 simulations out of 10, however, the ratio increases. A threshold might have then been hit, from which the Earth analog enters a regime where the cometary Xe signature is not overwritten by collisions with asteroids anymore. In these 3 cases, there is one single cometary impactor every $\simeq$ 1000 chondritic impactors. If we consider that the total mass of comets accreted to Earth ($9.76\times10^{21}$g) is equivalent to 0.03 wt.\% of the late accretion, taken as 0.5 wt.\% of the Earth \cite{Bekaert2020}, and that carbonaceous asteroids represent 30 wt.\% of the late accretion, one cometary impactor every 1000 carbonaceous impactors of the same size would have led to the current Xe isotopic signature in the Earth's atmosphere. 

These results are all the more valid given that our simulations underestimate the true contribution of carbonaceous material before the Moon-forming impact, and overestimate it after the Moon-forming impact (cf. Section \ref{CC_cont}).

We also highlight that there is no clear correlation between the comet/carbonaceous planetesimal ratio and the time of the Moon-forming impact. While the weighted cumulative collision rates of carbonaceous planetesimals remain constant during the late accretion (around $10^{-2}$ by carbonaceous planetesimals), it varies a lot for comets (between $10^{-5}$ and $10^{-9}$ by comet) and this variation is driven by a stochastic behavior \cite{Joiret2023}.

Comets in our simulations are assumed to have around the same mass as Pluto and there might have been around 1000 of these objects in the primordial Kuiper Belt \cite{Nesvorny2016}, so between $10^{-2}$ and $10^{-6}$ collisions of this type could have happened during late accretion. In other words, this is not very probable. Considering that the size frequency distribution of comets goes as $N(>d)$ $\simeq$ $d^{-q}$ , with $4.5 < q < 7.5$, for objects with a diameter d$>$100 km \cite{Fraser2014, Morby2021}, the number of objects larger than one third the size of Pluto, could be 140 to 3800 times more numerous than Pluto-mass objects. The cumulative collision rate between these objects and Earth would thus range between 1.4 and 38 collisions in the best-case scenarios. They could have played an important role in the cometary contribution to Earth.

\begin{table*}[t!] %[tb]%[!ht]
\begin{center}
\begin{tabular}{lp{1.5cm}p{1.5cm}p{1.5cm}p{2cm}p{2cm}p{2cm}p{2cm}p{2cm}}
    \hline \hline
          & Time of MFI (Myr) & Time of 63\% acc. $\tau$ (Myr) & MF impactor mass ($M_{\Earth}$) & Carb. mass wrt. Earth analog mass (wt.\%) & LA mass wrt. Earth analog mass (wt.\%) & Carbonaceous fraction of the LA (wt.\%) & Cometary fraction of the LA (wt.\%) \\
 \textbf{constraints} & \textbf{[30-100]} & & \textbf{$\simeq$ 0.1} & \textbf{$<$ 10} & \textbf{$\simeq$ 0.5} & \textbf{/} & \textbf{$\simeq$ 0.03} \\ \hline
 CL1 & 4.3 & 2.2 & 0.07 & 1.3 & 2.2 & 32.4 & 2.75 \\ 
 CL2 & 49.8 & 7.2 & 0.07 & 1.5 & 0.4 & 27.5 & 0.0006 \\ 
 CL6 & 40.8 & 8.5 & 0.27 & 1.3 & 0.3 & 36.4 & 0.0007 \\ 
 CL13 & 16.7 & 16.7 & 0.35 & 1.5 & 1.6 & 27.5 & 2.21 \\
 CL14 & 39.0 & 10.7 & 0.13 & 1.1 & 0.3 & 26.5 & 3.91 \\
 CL16 & 95.7 & 95.7 & 0.48 & 1.1 & 0.01 & 0.0 & 0.0 \\
 CL17 & 45.6 & 2.8 & 0.14 & 1.3 & 0.3 & 36.7 & 0.006 \\ 
 CL18 & 10.7 & 10.7 & 0.27 & 1.3 & 1.4 & 26.3 & 0.06 \\ 
 CL19 & 73.1 & 4.3 & 0.07 & 1.4 & 0.08 & 50.0 & 0.3 \\
 CL20 & 13.2 & 6.8 & 0.07 & 1.5 & 1.7 & 34.5 & 0.004 \\ \hline \hline
\end{tabular}
\caption{\small Summary of all successful simulations. The values specific to the solar system are displayed for comparison. Time of 63\% accretion refers to the time at which the Earth analog has accreted 63\% of its final mass. MFI, MF and LA stand for Moon-forming impact, Moon-forming and Late Accretion respectively.}
\label{table2}
\end{center}
\end{table*}

\section{Discussion}
\subsection{Comparison with previous studies}
\citet{Nesvorny2023} developed a dynamical model for impacts from comets, asteroids and leftover planetesimals in the inner solar system. They obtained the collision rates of asteroids directly from their integrator and those of comets from the {\"O}pik method, like we do in the present study. However, they started from a different distribution of planetesimals and cometesimals. In particular, they considered a distribution of smaller sized comets. They calculated the collision rates between comets (mainly comprised between 10 and 100 km) and a target body of 1 $M_\Earth$ at 1 AU, and obtained a smooth profile for the cometary bombardment after the Moon-forming impact. In this study and in \citet{Joiret2023}, we mainly focus on bigger objects and emphasize the importance of the stochastic component of the cometary bombardment. \citet{Nesvorny2023} found that leftover planetesimals (called locally-sourced planetesimals in our study) dominated the impacts on Earth after the Moon-forming impact. They showed that $\simeq$ 6 $\times$ $10^{4}$ $>$10 km leftover planetesimals, $\simeq$ 500 $>$10 km comets and $\simeq$ 100 $>$10 km asteroids impacted the Earth before the Moon-forming impact, and $\simeq$ 4 $\times$ $10^{4}$ $>$10 km leftover planetesimals, $\simeq$ 200 $>$10 km comets and $\simeq$ 100 $>$10 km asteroids after the Moon-forming impact (see Fig. 13 in \citet{Nesvorny2023}). According to their results, the cumulative collision rate of comets relative to asteroids or leftover planetesimals decreases after the Moon-forming impact. Therefore, they do not explain the Xe signature dichotomy between the Earth's mantle and atmosphere. 

\begin{figure}[h!]
  \centering
  \includegraphics[width=12cm]{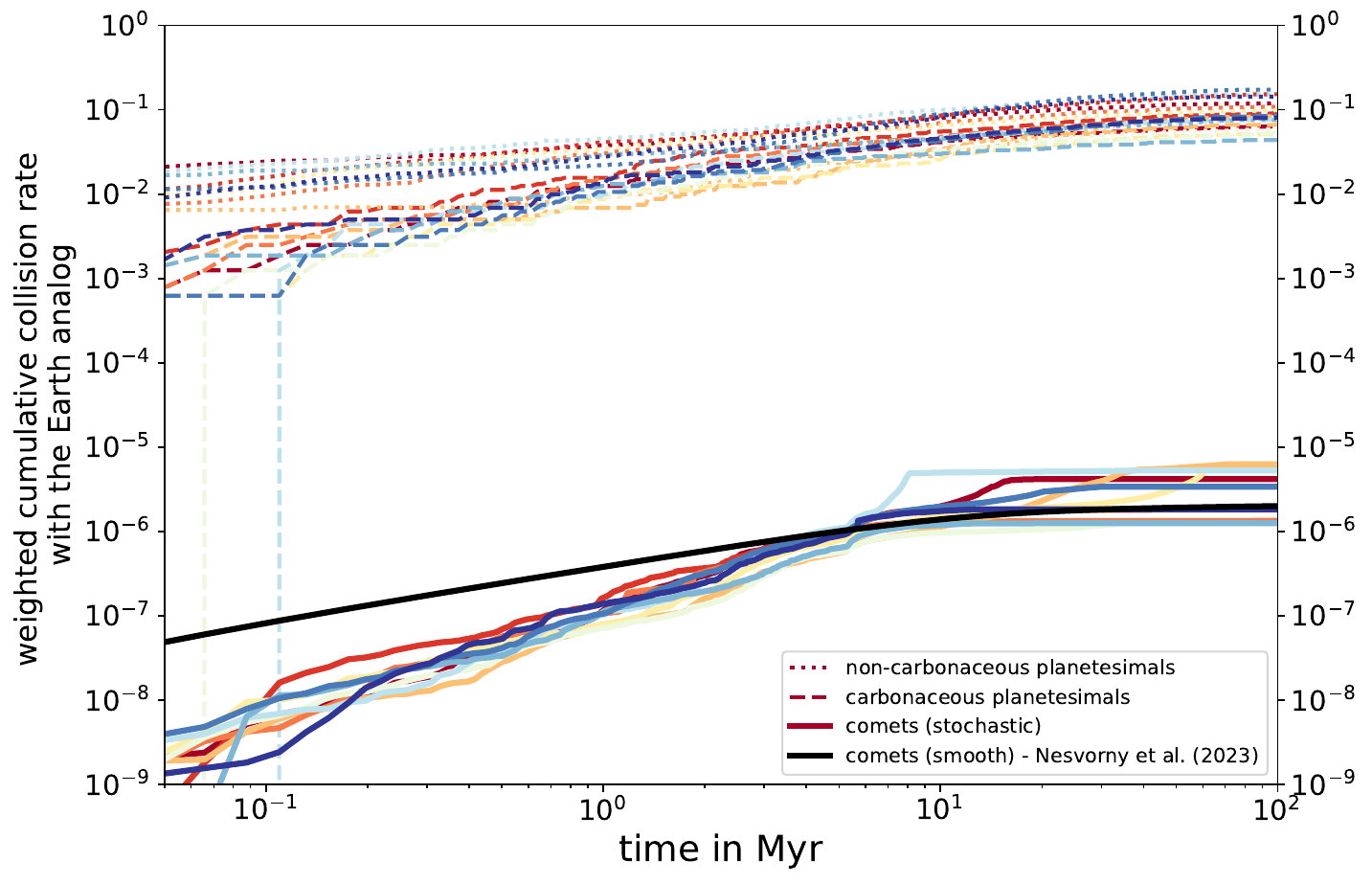}
  \caption{\small Comparison of the weighted average cumulative collision rate of a non-carbonaceous (locally-sourced) planetesimal (dotted lines), a carbonaceous planetesimal (dashed lines), and a comet (solid lines) respectively, with an Earth analog. The black solid line shows the contribution of smaller comets, as obtained by \citet{Nesvorny2023}, while the coloured solid ones show the contribution of larger comets (this study).}
  \label{comp_Nes}
\end{figure}

We include the cometary influx curve obtained by \citet{Nesvorny2023} to Figure \ref{comp_ascom}. In order to compare similar methodologies, we reassess our collision probability calculations, only taking into account the main Earth analog embryo - i.e. with a growing radius and located around 1 AU - and neglecting the other embryos that will form the Earth. The results are shown in Figure \ref{comp_Nes}. It appears that the smooth bombardment dominates until t $\simeq$ 4 Myr and the stochastic influx takes over thereafter. However, using a target body of 1 $M_\Earth$ at 1 AU to calculate impact probabilities, as it is done in \citet{Nesvorny2023}, slightly overestimates the cometary bombardment in the first millions years. Indeed, collision rates directly depend on the cross sections of the colliding particles $(R + r)^{2}$, with R the radius of the Earth and r the radius of the comet. Hence the method we use in this paper, considering the smaller radii of the Earth analog embryos at the beginning of the simulations, and taking them all into account, is more accurate. In fact, \citet{Nesvorny2023} emphasize that their curve for cometary bombardment should not be used for t $\simeq$ 0 Myr, as the main focus of their study is the late accretion.

In their study, \citet{Dauphas2002} argue that comets represented less than $10^{-3}$ by mass of the late impacting population, basing their reasoning on the geochemistry of noble metals and gases. Similarly, \citet{Bekaert2020} showed that the total mass of comets accreted by Earth is equivalent to 0.03 wt.\% of the late accretion. In other words, about 1 in 3000 late impactors of the same size would be a comet. Considering that carbonaceous asteroids represent 30 wt.\% of the late accretion (cf. Fig. \ref{carb_LV}), it would amount to about 1 comet every 1000 carbonaceous impactor of the same size. These geochemical results are perfectly consistent with 30\% of our simulations (cf. Fig \ref{comp_ascom_MFI}).

\subsection{What if the instability happened later?}
\citet{Morby2018} and \citet{Mojzsis2019} both argued that the instability happened within the first 100 Myr of solar system history, using observational and geochemical evidence respectively. In this context, most of the cometary bombardment might have taken place after the Moon-forming impact.  

Taking into account an early giant planet instability, we find a roughly constant ratio of $\simeq$ $10^{-4}$ between the cumulative collision rates of comets and those of carbonaceous planetesimals before the Moon-forming impact, then a large dispersion of values afterward (cf. Fig \ref{comp_ascom_MFI}). If the instability happened later, the cometary bombardment might have happened mainly \textit{after} the Moon-forming impact and the ratio would be much lower \textit{before}. However, after the Moon-forming impact, the ratio would probably tend to even higher values. Therefore, a later instability would help explain the Xe isotopic dichotomy between the Earth's atmosphere and mantle.

In addition to this, we do not expect that a late instability would affect the contribution of carbonaceous planetesimals to the Earth's mantle. The principal contribution of carbonaceous material to the Earth's mantle might have occurred predominantly during the giant planets' growth \cite{Raymond2017}.

\subsection{Limitations}
Our findings are primarily limited by the fact that the isotopic composition of noble gases in comets is exclusively inferred from the Rosetta measurements on 67P/C-G. While it cannot be established that all cometary bodies have the same noble gas isotopic signature as 67P/C-G, laboratory measurements showed that volatile content in cometary ice reflects gas pressure and temperature under which they formed \cite{BarNun1985, Yokochi2012, Almayrac2022}. Hence, provided that all comets formed in the same region, and that they did not lose a significant portion of their volatiles over their thermal histories \cite{Gkotsinas2022}, their volatile content should be similar to that of 67P/C-G \cite{Marty2016}. Additionally, the very unique isotopic signature of primordial Xe in our atmosphere being similar to the Xe signature of 67P/C-G would be hard to explain if the latter signature is not representative of a significant subset of the cometary reservoir.  

Another important limitation is our assumption that there is no loss of volatile elements to space during collisions between comets or carbonaceous chondrites and terrestrial embryos. In the case of the Earth, this assumption was shown to be roughly correct for collisions occurring after the Moon-forming impact \cite{Marty2016} but it becomes less valid in the case of giant impacts which likely involve significant atmospheric loss \cite{Genda2005, Kegerreis2020}. While \citet{deNiem2012} have shown that single impacts of very massive bodies are of more importance for the atmospheric evolution than other input parameters, \citet{Sinclair2020} argue that only the smallest cometary impactors could contribute material to the atmosphere. However, these studies often consider atmospheric conditions that are analogous to those of our current atmosphere. Yet, during the giant impact phase, the Earth's atmosphere was oftentimes in a runaway greenhouse state \cite{Abe1985, Abe1986, Zahnle1988} and was thus very thick and dense. After the mantle solidified, the steam atmosphere condensed to form a warm ($\simeq$ 500 K) water ocean under a thick ($\simeq$ 100 bar) $CO_{2}$-atmosphere \cite{Zahnle2007}. $CO_{2}$ is assumed to subduct in the mantle on a 10 - 100 Myr timescale \cite{Zahnle2007}. Under these pressure and temperature conditions, fragmentation and ablation of big impactors become critical. \citet{Shuvalov2014} showed that impacts of 1 - 10 km sized projectiles in hot and dense atmospheres lead to aerial bursts which in turn result in strong atmospheric erosion. The amount of accreted material is also very high for these type of impacts \cite{Shuvalov2014, Kral2018}. Hence, big impactors, and in particular big comets, probably strongly influenced the evolution of our early atmosphere. 

Finally, the computational cost of our simulations imposes a limit on their resolution. Indeed, one simulation takes about a month on one GPU, for a 100 Myr integration. The outer disk of comets is comprised of 10000 Pluto-mass objects to account for the 25 $M_\Earth$ expected in this region initially. These starting conditions are not too unrealistic, since it is expected that there might have been $\simeq$ 1000 Pluto-mass objects in the primordial Kuiper belt \cite{Nesvorny2016}. So our simulations mostly reflect on a stochastic component of the cometary bombardment, dictated by these Pluto-mass objects. This also has repercussions on the calculated collision rates of comets, which could be slightly overestimated given that they directly depend on the cross section of the colliding particles $(R + r)^{2}$. Nevertheless, this is not too significant, as the radius of the Earth analog R is the dominant term. In addition, we also show in Section \ref{validation} that the {\"O}pik/Wetherill algorithm we use to obtain the collision rates of comets might underestimate it by a factor of two. Therefore, these effects may offset each other.

\section{Summary and Conclusions}
We have performed dynamical simulations of the early stages of solar system evolution (i.e. the first 100 Myr after gas disk dispersal) in the context of an early giant planet instability. We have included terrestrial embryos and planetesimals, carbonaceous planetesimals, giant planets, and comets; and have assessed the collision rates between both carbonaceous planetesimals and comets and the Earth constituent embryos/planetesimals. Our main results can be summarized as follows:
\begin{itemize}
     \item Cometary isotopic signatures are essentially detectable on a planetary body through noble gases (in particular Ar, Kr and Xe), because they are 3 to 4 orders of magnitude more abundant in comets than in carbonaceous chondrites \cite{Bekaert2020}. The abundance of $H_{2}O$, C and N on the other hand, are only about 1 order of magnitude more abundant in comets than in carbonaceous chondrites \cite{Bekaert2020}. Our simulations show that there are about 1 comet every $10^{4}$ carbonaceous asteroids colliding with the Earth before the Moon-forming event. Consequently, the very high collision rates of carbonaceous asteroids relative to comets on Earth may have balanced out the high abundance of volatiles, including noble gases, in comets or may have permitted up to 10\% cometary Xe at most in the Earth's mantle. These results concur with geochemical studies which have shown that the Xe isotopic signature in the Earth's mantle is chondritic \cite{Peron2018, Broadley2020}, all the more so since measurement uncertainties do not preclude a small contribution from comets.

    \item After the Moon-forming event and magma ocean solidification, volatile-rich bodies impacting the Earth mostly contributed to the atmosphere's volatile budget. Our simulations show that after this Moon-forming impact, there is a non-negligible probability (3 simulations out of 10) that the number of carbonaceous asteroids relative to the number of comets colliding with the Earth decreases. This decrease in relative contribution is crucial because it might reach a point where the cometary isotopic signature is not overwritten by collisions with asteroids anymore. In these simulations, there is about 1 comet every $10^{3}$ carbonaceous asteroids colliding with the Earth \textit{after} the Moon-forming event (in contrast to 1 comet every $10^{4}$ carbonaceous asteroids \textit{before}). Given the noble gases concentrations of comets relative to those of carbonaceous asteroids, such a scenario is fully consistent with the geochemical measurements that constrain the cometary volatiles contribution on Earth to $\le$ 1\% for $H_{2}O$, C and N, but a few tens of \% for noble gases in the atmosphere \cite{Marty2012}. These constraints also require that the total mass of comets accreted by Earth after the Moon-forming impact is equivalent to 0.03 wt.\% of the late accretion \cite{Bekaert2020}, corresponding to 1 comet every $10^{3}$ carbonaceous asteroids - as carbonaceous asteroids could represent $\simeq$ 30 wt.\% of the late accretion. This is in excellent agreement with our findings. More importantly, this scenario is compatible with and provide an explanation to the Xe isotopic dichotomy between the Earth's mantle and atmosphere \cite{Marty2017, Peron2018, Broadley2020}. It is important to note that what increases after the Moon-forming impact is not the probability of collision with a comet, but the ratio of comets to carbonaceous asteroids brought to Earth. This happens in 30\% of our simulations because the cumulative collision rate (or number of collisions) of carbonaceous asteroids with Earth \textit{after} the Moon-forming impact is always lower than \textit{before}, while the much smaller cumulative collision rate of comets to Earth might remain the same - due to the stochastic component of the cometary bombardment.

    \item Based on our simulations and initial conditions, we find that carbonaceous asteroids accounted for 1 to 2 wt.\% of the Earth's final mass. This result is consistent with the constraint that the Earth's highly volatile budget requires the accretion of $\simeq$ 2 ($\pm$1) wt.\% of carbonaceous material \cite{Marty2012}. However, it is somewhat lower than more recent studies that limit the carbonaceous contribution to Earth to $\simeq$ 4 wt.\% \cite{Burkhardt2021} or $\simeq$ 6 wt.\% for CI-like material \cite{Savage2022}, or 10 wt.\% at most for CV-like material \cite{Kleine2023}. 
    
    Starting from the same set of initial conditions, we find that $\simeq$ 30\% of the late accretion is comprised of carbonaceous planetesimals, with the remaining mass being non-carbonaceous. These simulation outcomes are not sensitive to the time of the Moon-forming impact, and they match bulk silicate Earth's Mo and Zn isotopic measurements. As moderately siderophile and slightly siderophile elements respectively, Mo and Zn are expected to have been delivered late \cite{Dauphas2017, Mahan2018}. It is estimated that $\simeq$ 30\% of Earth's Mo \cite{Burkhardt2021} and $\simeq$ 1/3 of Earth's Zn \cite{Sossi2018, Savage2022} was delivered by carbonaceous planetesimals. However, our results diverge from the constraints indicating that, if the impactites are representative of late accretion in general, only up to $\simeq$ 20 to 25\% of carbonaceous material - and potentially much less - contributed to the late accretion \cite{Worsham2021}.
    
    Our model thus fails to explain a total carbonaceous proportion higher than 1-2 \%, while being consistent with a non-carbonaceous-dominated late accretion. New approaches should be investigated to address these combined constraints. One hypothesis is that a significant proportion of the Earth's carbonaceous budget was accreted before gas disk dispersal. An early accretion is indeed expected due to important ejections of carbonaceous planetesimals to the inner solar system as Jupiter and Saturn were growing \cite{Raymond2017}. Assuming this is true, it would further support our dynamical interpretation for the xenon signature dichotomy, as the ratio of cometary to carbonaceous material would increase even more at later times.

    \item We infer the time of the Moon-forming impact based on the late accretion mass obtained in our simulations, using the method of \citet{Jacobson}. Provided that the late accretion mass is equivalent to the \textit{late veneer} mass extrapolated from HSE abundances in the mantle, we find that the Moon-forming event happened between $\simeq$ 24 and 67 Myr after gas disk dispersal. Assuming a gas disk lifetime of 4 Myr \cite{Haisch2001, Wang2017}, this is equivalent to between $\simeq$ 28 and 71 Myr after the condensation of the first solids in the solar system. However, this time range contains large uncertainties associated with the asteroids size frequency or the abundance of HSEs accreted before the Moon-forming impact. 

\end{itemize}

\section{Acknowledgements}
Computer time for this study was partly provided by the computing facilities MCIA (Mésocentre de Calcul Intensif Aquitain) of the Université de Bordeaux and of the Université de Pau et des Pays de l'Adour, France. Numerical computations were also partly performed on the S-CAPAD/DANTE platform, IPGP, France. We acknowledge support from the CNRS MITI funding program (PhD grant to S.J., NOBLE project) and the CNRS's Programme National de Planétologie (PNP). We acknowledge funding in the framework of the Investments for the Future programme IdEx, Université de Bordeaux/RRI ORIGINS. This project has received funding from the European Research Council (ERC) under the European Union’s Horizon Europe research and innovation program (grant agreement no. 101041122 to G.A.). M.S.C. is supported by NASA Emerging Worlds grant 80NSSC23K0868 and NASA’s CHAMPs team, supported by NASA under Grant No. 80NSSC21K0905 issued through the Interdisciplinary Consortia for Astrobiology Research (ICAR) program.  

\medskip
\bibliography{reference2}
\end{document}